\documentclass{IEEE_lsens}

\usepackage{textcomp}
\usepackage[noadjust]{cite}
\ifCLASSINFOpdf
\else
\fi
\usepackage[T1]{fontenc}
\usepackage{amsmath}
\usepackage[cmintegrals]{newtxmath}
\usepackage{bm}
\usepackage{orcidlink}
\usepackage{array}
\usepackage{url}
\usepackage{graphicx}
\usepackage{multirow}
\usepackage{subfigure}
\usepackage{adjustbox} 
\usepackage{array} 

\ifCLASSINFOpdf
\else
\fi
\providecommand{\hypersetup}[1]{\relax}

\hypersetup{pdftitle={Bare Demo of IEEE\_lsens.cls for IEEE Sensors Letters},
pdfsubject={Typesetting},
pdfauthor={Michael D. Shell},
pdfkeywords={Class, IEEE, IEEE\_lsens, IEEE Sensors Letters, LaTeX, Typesetting, TeX}}
\begin{document}
\IEEELSENSarticlesubject{Sensor Signal Processing}

%
\title{ILD-VIT: A Unified Vision Transformer Architecture for Detection of Interstitial Lung Disease from Respiratory Sounds}

\author{
Soubhagya Ranjan Hota$^\dag$, Arka Roy\IEEEauthorieeemembermark{1}$^\dag$, and Udit Satija\IEEEauthorieeemembermark{2}\\
Department of Electrical Engineering, Indian Institute of Technology Patna, Bihar, 801106, India\\
$^\dag$ Contributed equally and can be considered as first authors
\IEEEauthorblockA{
\IEEEauthorieeemembermark{1}Graduate Student Member, IEEE,
\IEEEauthorieeemembermark{2}Senior Member, IEEE
}
\thanks{Corresponding author: Udit Satija (e-mail: \text{udit@iitp.ac.in}).\protect\\}
}


\IEEEtitleabstractindextext{%
\begin{abstract}
Interstitial lung disease (ILD) represents a group of restrictive chronic pulmonary diseases that impair oxygen acquisition by causing irreversible changes in the lungs such as fibrosis, scarring of parenchyma, etc. ILD conditions are often diagnosed by various clinical modalities such as spirometry, high-resolution lung imaging techniques, crackling respiratory sounds (RSs), etc. In this letter, we develop a novel vision transformer (VIT)-based deep learning framework namely, ILD-VIT, to detect the ILD condition using the RS recordings. The proposed framework comprises three major stages: pre-processing, mel spectrogram extraction, and classification using the proposed VIT architecture using the mel spectrogram image patches. Experimental results using the publicly available BRACETS and KAUH databases show that our proposed ILD-VIT achieves an accuracy, sensitivity, and specificity of 84.86\%, 82.67\%, and 86.91\%, respectively, for subject-independent blind testing. The successful onboard implantation of the proposed framework on a Raspberry-pi-4 microcontroller indicates its potential as a standalone clinical system for ILD screening in a real clinical scenario.
\end{abstract}

\begin{IEEEkeywords}
Respiratory sounds, interstitial lung disease (ILD), vision transformer, disease detection.
\end{IEEEkeywords}}

\maketitle

\section{Introduction}
Interstitial lung disease (ILD) is a general term for a group of chronic respiratory conditions developed by lung tissue scarring and fibrosis, which stiffens the lungs and makes it difficult to breathe and acquire enough oxygen \cite{7422082}. ILD can arise due to genetic abnormalities, autoimmune diseases including sarcoidosis or rheumatoid arthritis, exposure to toxic pollutants, etc \cite{BTS1999}. A recent study from Lancet journal reported that from 2005-2020, the overall prevalence of ILD increased from $24.7$\% to $33.6$\% \cite{Travis2013}. As the lung undergoes though irreversible changes due to ILD, it is essential to detect ILD at an early stage to provide the necessary medication to control the disease progression. The standard techniques to identify ILD involve X-rays \cite{BTS1999}, Computed Tomography(CT) scans \cite{7422082}, high-resolution CT (HRCT) scan \cite{9725491}, spirometry tests or lung functions test \cite{BTS1999}, etc. However, the majority of these modalities are either costly or require high patient effort, and skilled technicians \cite{vanderBruggenBogaarts1995}. In contrast, chest auscultation plays a major role for the physician in identifying abnormalities by observing various adventitious respiratory sounds (RSs), such as wheeze, crackle, sqwaks, etc. \cite{arka_tasl, 9881964}, that are produced due to the presence of disease. Therefore, exploiting RSs in conjunction with artificial intelligence-based automated algorithms will be beneficial for the screening of ILD.

The majority of the works on ILD are related with the analysis of CT or HRCT images. Anthimopoulos et al. \cite{7422082} used CT images with five-layer CNN architecture and achieved an accuracy of 85.5\%.  Martinez et al. \cite{Martinez} also used CT images with an ensemble-based approach and got an accuracy of 82.7\%. Vishraj et al. \cite{9725491} exploited Haralick features from the HRCT images and fed them to a random forest (RF) based machine learning classifier. By using this approach an accuracy of 85.8\% was achieved. In recent years Roy et al. \cite{arka_ild} for the first time exploited RSs for ILD detection via sinc-convolution-based 1D-deep learning (DL) architecture and achieved accuracy, sensitivity, and specificity as 81.25\%, 78.85\%, and 83.33\%, respectively.

The major contributions of this letter are: (I) here, we first investigated the potential of mel spectrogram time-frequency representation (TFR) of RSs in conjunction with a vision transformer (VIT)-based DL architecture for ILD detection; (II) the proposed framework is extensively evaluated on two publicly available databases using various performance measures, (III) Analysis of impact of various noises on the classification at different SNR levels, and lastly (IV) implemented on a low compute microcontroller: Raspberry Pi-4, for the first time, which shows the potential of being translated in real-clinical scenario. The rest of the letter is organized as: Section \ref{sec: dataset} provides a brief discussion about the public databases, Section \ref{sec: method} describes the proposed framework in detail, Section \ref{sec: result} evaluates the framework, and Section \ref{sec: colclusion} concludes the research. 

\section{Database Description}
\label{sec: dataset}
In this letter, we have utilized two publicly available databases: (a) BRACETS \cite{pessoa2023bracets} and (b) KAUH \cite{FRAIWAN2021106913}. The BRACETS database consists of simultaneous recordings of electrical impedance tomography (EIT) and RSs from subjects with asthma, ILD, COPD, and healthy groups \cite{pessoa2023bracets}. The RSs of both databases were recorded using 3M Littmann 3200 digital stethoscope with a sampling rate of 4000 Hz. In this work, we have used a total of 384 and 176 RS recordings from 17 ILD-affected and 8 healthy subjects from the BRACET database. Due to the limited availability of healthy recordings, we have curated 115 RS recordings from 30 healthy subjects from the KAUH database. The length of the RSs varies erratically in both databases from 15 to 50 sec. The detailed demographic information is available in \cite{pessoa2023bracets,FRAIWAN2021106913}.
\vspace{-0.5cm}
\section{Proposed Framework}
\label{sec: method}
\begin{figure}[htbp!]
    \centering
    \includegraphics[width=0.5\textwidth, height=4cm]{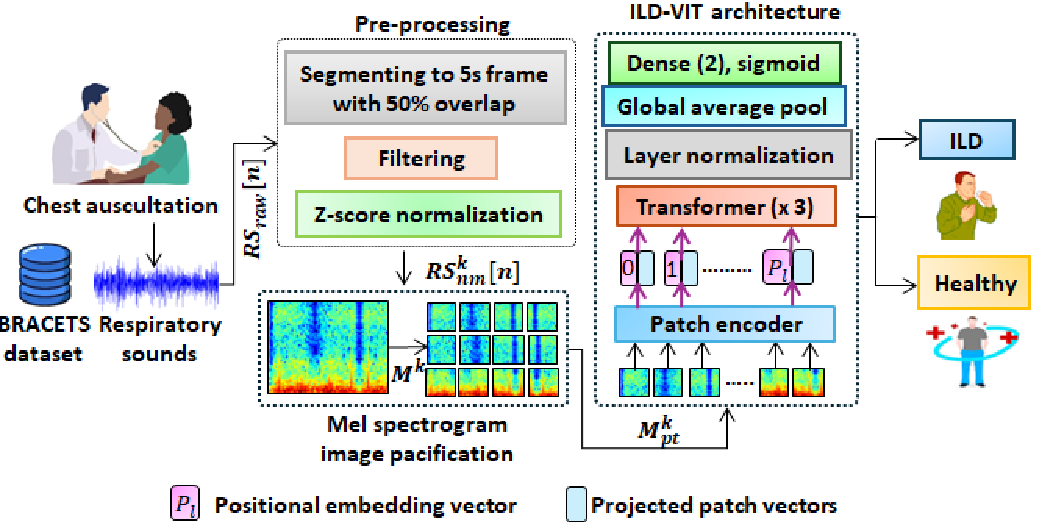}
    \vspace{-0.35cm}
    \caption{Illustrates the block diagram of the proposed ILD-VIT framework for RS-based ILD detection.}
    \vspace{-0.35cm}
\label{fig:Block_diagram}
\end{figure}

In this section, we discuss the proposed framework for ILD detection. The proposed framework, as shown in Fig. \ref{fig:Block_diagram} contains three stages: (a) pre-processing, (b) mel-spectrogram patch extraction, (c) ILD detection using our proposed vision transformer architecture. The following subsection details the function of each stage.
\subsection{Pre-processing}
\begin{figure}
     \centering
     \includegraphics[width=0.5\textwidth]{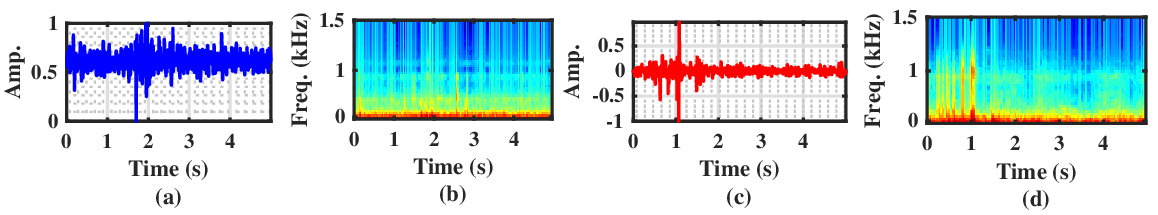}
     \vspace{-0.6cm}
     \caption{Illustrates the temporal RSs and their mel spectrogram TFRs for (a-b) healthy and (c-d) ILD cases, respectively.}
     \vspace{-0.4cm}
     \label{fig:ild_mel}
 \end{figure}

Initially, the raw RSs ($RS_\text{raw}[n]$) are segmented into 5 sec frames $(W_f)$ with the 50\% overlap $(O_\text{v})$, which can be described as \cite{10288401}:
\begin{equation}
    RS_{seg}^{k}[n]=RS_{raw}\left[W_f \cdot k \cdot\left(1-O_{v}\right)+n\right], \hspace{0.3cm} k=0,1,2, \ldots K
\end{equation}
where $RS_{seg}^k[n],~K$ denote the $k^{th}$ segmented frame RS and total number of frames respectively. Thereafter, these segmented frames $(RS_{seg}^{k}[n])$ are passed through a 4$^{th}$ order Butterworth high-pass filter (HPF) with a cut-off frequency of 10 Hz and subsequently normalized via z-score normalization as:  $RS_{nm}^k[n]=\frac{RS_{flt} ^k[n]-\mu_{flt}}{\sigma_{flt}}$, where $RS_{flt} ^k[n],~RS_{nm}^k[n], ~\mu_{flt},$ and $\sigma_{flt}$ denote the $k^{th}$ filtered, normalized frame of RS, mean and stand deviation of the filtered RS signal. Fig. \ref{fig:ild_mel}(a) and  Fig. \ref{fig:ild_mel}(c) illustrate the normalized time-domain RS frame of healthy and ILD-affected subjects. After segmenting the entire (combined dataset), a total of 1691 and 1962 segments were obtained for healthy and ILD classes.

\subsection{Mel spectrogram patch extraction}
As RSs are highly non-stationary, we transform the time signals to mel spectrogram time-frequency response (TFR) which captures the time-varying frequency content of RSs effectively \cite{9756949}. 
To extract mel spectrogram from $RS_{nm}^k[n]$ first the short time Fourier transform $(S^k[g, f])$ is evaluated as \cite{arka_tasl}:

\begin{equation}
    S^k[g, f]=\sum_{n=0}^{N-1} RS_\text{nm}^k[n] \cdot \omega[n-g H] \cdot e^{-j \frac{2 \pi n f}{N}}
\end{equation}
where $\omega[n]$ is a Hanning window of 1024 samples with a hop-length $(H)$ of 512 samples. Then the Hertz frequencies $(f)$ are mapped to mel-scale frequency $(f_{mel})$ to create the triangular mel filter banks as \cite{arka_memea}: $f_{mel}= 2595\cdot \log (1+ f/700)$.
Finally, the mel spectrogram ($M^k[g,f]$) is generated by multiplying the magnitude of $S^k[g, f]$ with the mel-filter banks. In this letter, we have considered a total of $64$ mel filter. Later, these 2D $M^k[g,f]$ are converted to 3-channel images via `jet' color map \cite{arka_memea} to produce a data size of $64 \times 64 \times 3$. Fig. \ref{fig:ild_mel}(b) and  Fig. \ref{fig:ild_mel}(d) indicate the mel spectrograms for healthy and ILD-affected subjects. Later, these ${M}^k $ are split into 2D flattened patches: $M_{pt}^{k} \in \mathbb{R}^{N_{pt} \times 3Pt^2}$, where $(P_t,~P_t)$ is the spatial resolution of patch and $N_{pt}$ is the number of such flattened patches, which can be expressed as \cite{dosovitskiy2020vit,10376240}:  For $N_{pt} = \frac{64 \times 64}{Pt^2}$.
\subsection{Architectural details ILD-VIT}
In this subsection, we discuss the proposed ILD-VIT architecture for ILD detection based on mel spectrogram patches $(M_\text{pt}^{k})$ of RSs which are first projected through a linear layer with embedding with projection length $(P_{ln})$ of 64 which is decried as \cite{dosovitskiy2020vit}:
\begin{equation}
    \mathbf{Y^k}= (M_{pt}^{k}\cdot \mathbf{W}_{ptj}+ \mathbf{B}_{ptj})+\mathbf{W}_{pos},
\end{equation}
Here, $\mathbf{Y^k}\in \mathbb{R}^{N_{pt}\times P_{ln}}, ~\mathbf{W}_{ptj} \in \mathbb{R}^{3Pt^{2}\times N_{pt}}, ~\mathbf{B}_{ptj}\in \mathbb{R}^{P_{ln}}, ~\mathbf{W}_{pos}\in \mathbb{R}^{N_{pt}\times P_\text{ln}}$ represent the patch encoder layer's output, weight, and positional embedding matrix, respectively.

\begin{figure}
    \centering
    \includegraphics[width=0.5\textwidth]{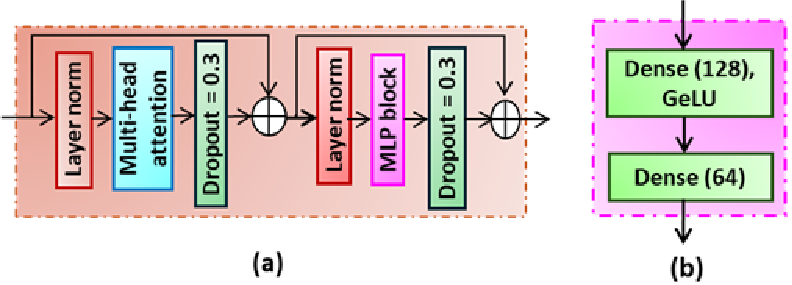}
    \vspace{-0.5cm}
    \caption{Internal architecture of (a) transformer and (b) MLP block.}
    \vspace{-0.45cm}
    \label{fig: mha_ild}
\end{figure}

Thereafter, these patch-encoded data are passed through three transformer blocks which comprise a multi-head attention layer (MHA) and multi-layer perceptron (MLP) block, where layer normalization (LN) is added prior to MHA, MLP layer and dropout layer ($D_p$) is added after each MHA, MLP block, followed by an identity skip connection as shown in Fig. \ref{fig: mha_ild}. The MLP block comprises a dense layer with 128, 64 neurons ($Dn_{k}$) and GeLU activation. Each MHA block contains $N_{H}$ parallel heads with separate learnable self-attention parameters: query $(\mathcal{Q})$, value $(\mathcal{V})$, and key $(\mathcal{K})$, which are derived from $\mathbf{Y^k}$. The mathematical description of the MHA block is given as \cite{dosovitskiy2020vit,10376240}:
\begin{subequations}
\begin{equation}
    f^{SA}_{i}(\mathbf{Y^k})=\operatorname{Softmax}(\mathcal{Q}_{i}\mathcal{K}_{i}^{T}/\sqrt{Dn_k})\mathcal{V}_{i}, \hspace{0.3cm} i=1,2,..N_{H}
\end{equation}
\begin{equation}
    f^{MHA}(\mathbf{Y^k})=[f^{SA}_{1}(\mathbf{Y^k}),..., f^{SA}_{4}(\mathbf{Y^K})]\mathbf{W}_{MHA}+\mathbf{B}_{MHA},
\end{equation}
\end{subequations}

Where $f^{SA}(\cdot), ~f^{MHA}(\cdot), ~\mathbf{W}_{MHA}, ~\mathbf{B}_{MHA}$ indicate the functional representation of the self-attention layer, MHA layer, the weight and bias matrix of the MHA layer respectively. Here $t^{th}$ transformer block's output can be represented as : 
\begin{subequations}
\begin{equation}
   \mathbf{\Tilde{Y}^{k}_{t}}=f^{DP}\left(f^{MHA}\left(f^{LN}\left(\mathbf{\Tilde{Y}^{k}_{t-1}} \right)\right)\right)+\mathbf{\Tilde{Y}^{k}_{t-1}}, \hspace{0.25cm} t=1,2,3
\end{equation}
\begin{equation}
    \mathbf{{Y}^{k}_{t}}=f^{DP}\left(f^{MLP}\left(f^{LN}\left(\mathbf{\Tilde{Y}^{k}_{t}}\right)\right)\right)+\mathbf{\Tilde{Y}^{k}_{t}}, \hspace{0.5cm} t=1,2,3
\end{equation}
\end{subequations}
where $f^{DP}(\cdot), ~f^{LN}(\cdot)$, and $f^{MLP}(\cdot)$ denote the functional representation of dropout, LN layer, and MLP block, respectively. Thereafter, the transformer encoded data is passed through another LN layer followed by a global average pooling (GAP) layer to create a 1D embedding of size $64\times 1$, which is finally classified into either Healthy or ILD, through a sigmoid-activated dense layer with 2 neurons. Table \ref{tab:proposed_model} illustrates the total parameter size of the proposed ILD-VIT architecture. Finally, the model is trained for 200 epochs via minimizing the binary cross entropy loss with a mini-batch gradient descent-based weight updation technique with a learning rate, batch size of 0.001, 64, respectively. 

\begin{table}[]
\centering
\huge
\caption{Parameter Size of the Proposed ILD-VIT
Architecture}
\label{tab:proposed_model}
\resizebox{\columnwidth}{!}{%
\begin{tabular}{|ccc|c|}
\hline
\multicolumn{1}{|c|}{\textbf{Layer type}} &
  \multicolumn{1}{c|}{\textbf{Specifications}} &
  \textbf{Dimension} &
  \textbf{Pparameters} \\ \hline
\multicolumn{1}{|c|}{\begin{tabular}[c]{@{}c@{}}Mel spectrogram\end{tabular}} &
  \multicolumn{1}{c|}{---} &
  $64\times64\times3$ &
  0 \\ \hline
\multicolumn{1}{|c|}{Patchification} &
  \multicolumn{1}{c|}{\begin{tabular}[c]{@{}c@{}}$P_t=8$\\  ${N_{pt}}=64$\end{tabular}} &
  $64\times192$ &
  0 \\ \hline
\multicolumn{1}{|c|}{\begin{tabular}[c]{@{}c@{}}Patchencoder\end{tabular}} &
  \multicolumn{1}{c|}{\begin{tabular}[c]{@{}c@{}}$P_{ln} = 64$\end{tabular}} &
  $64\times64$ &
  16448 \\ \hline
\multicolumn{1}{|c|}{\begin{tabular}[c]{@{}c@{}}Transformer module 1\end{tabular}} &
  \multicolumn{1}{c|}{\multirow{4}{*}{\begin{tabular}[c]{@{}c@{}}{$H_n$}=3,\\ $Dn_{k}=128,~64$,\\  ${D_p}=0.3$, $GeLU$\end{tabular}}} &
  $64\times64$ &
  83200 \\ \cline{1-1} \cline{3-4} 
\multicolumn{1}{|c|}{\begin{tabular}[c]{@{}c@{}}Transformer module 2\end{tabular}} &
  \multicolumn{1}{c|}{} &
  $64\times64$ &
  83200 \\ \cline{1-1} \cline{3-4} 
\multicolumn{1}{|c|}{\multirow{2}{*}{\begin{tabular}[c]{@{}c@{}}Transformer module 3\end{tabular}}} &
  \multicolumn{1}{c|}{} &
  \multirow{2}{*}{$64\times64$} &
  \multirow{2}{*}{83200} \\
\multicolumn{1}{|c|}{} &
  \multicolumn{1}{c|}{} &
   &
   \\ \hline
\multicolumn{1}{|c|}{LN} &
  \multicolumn{1}{c|}{---} &
  $64\times64$ &
  128 \\ \hline
\multicolumn{1}{|c|}{\begin{tabular}[c]{@{}c@{}}GAP\end{tabular}} &
  \multicolumn{1}{c|}{---} &
  $64\times1$ &
  0 \\ \hline
\multicolumn{1}{|c|}{Dense} &
  \multicolumn{1}{c|}{\begin{tabular}[c]{@{}c@{}}$Dn_{k}= 2$,  Sigmoid\end{tabular}} &
  $2\times1$ &
  130 \\ \hline
\multicolumn{3}{|c|}{\textbf{Total trainable parameters}} &
  \textbf{349506} \\ \hline
\end{tabular}%
} \vspace{-0.35cm}
\end{table}

\section{Result and discussion}
\label{sec: result}
In this section, we investigate the qualitative and quantitative performance of the proposed framework based on various performance measure and compare with noteworthy prior works on ILD detection.
\subsection{Performance evaluation}
 We used a wide range of performance metrics \cite{arka_tasl, 10453615, 9729415}, such as accuracy $(ACC)$, recall $(RCL)$ or sensitivity $(SNS)$, specificity $(SPF)$, precision $(PRE)$, ICBHI score $(IS)$, and F1-score $(FS)$. 
The equations for the aforementioned metrics are given as:
$ACC = \frac{Tr_p + Tr_n}{Tr_p + Tr_n + Fl_p + Fl_n}$, $ RCL =\frac{Tr_p}{Tr_p + Fl_n}$, $PRE =\frac{Tr_p}{Tr_p + Fl_p}$, $SPF =\frac{Tr_n}{Tr_n + Fl_p}$, $IS =\frac{SNS + SPF}{2}$, $FS =\frac{2\times PRE\times RCL}{PRE+RCL}$
where $Tr_p$, $Tr_n$, $Fl_p$, $Fl_n$ indicate the true positive, true negative, false positive, and false negatives, obtained from the confusion matrix while testing the proposed ILD-VIT.
\begin{figure}
    \centering
    \includegraphics[width=0.5\textwidth, height=2.5cm]{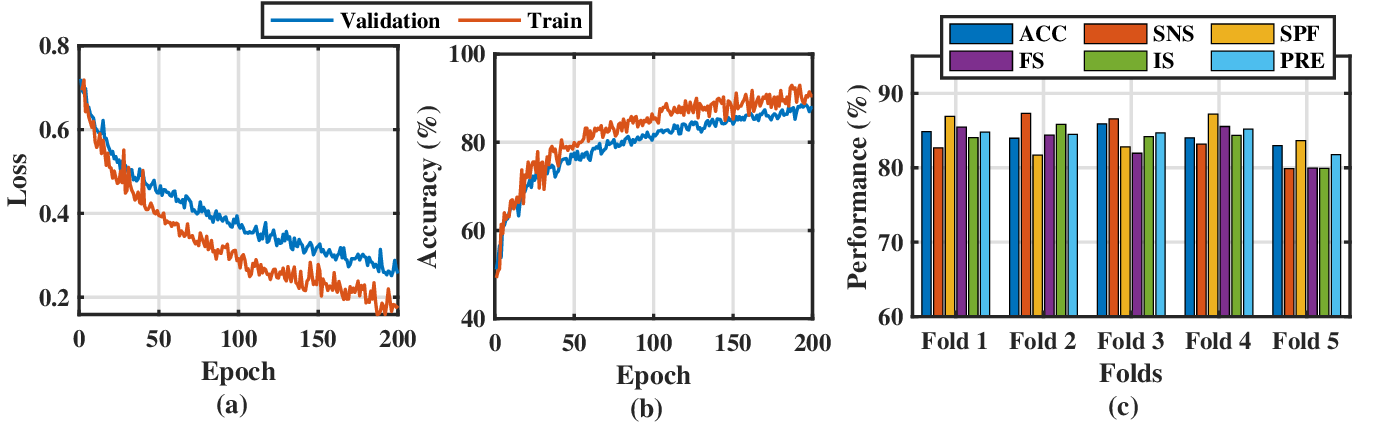}
    \vspace{-0.65cm}
    \caption{Illustrates (a-b) training-validation loss-accuracy curves for \textbf{Exp-1}, (c) fold-wise performance for \textbf{Exp-2}, obtained from our ILD-VIT}.
    \vspace{-0.35cm}
    \label{fig:final_loss_acc_curve}
\end{figure}

In this paper, we have done two experiments (Exp.): \textbf{Exp-1:} where we split the combined database into 70\%-10\%-20\% training-validation-testing sets on the subject level to ensure that signals from any subject do not appear in the train-validation-test sets, and thereafter performed the segmentation, TFR extraction, and subsequently trained the ILD-VIT architecture. From Fig. \ref{fig:final_loss_acc_curve}(a-b) shows the loss and accuracy curves of ILD-VIT model. From Fig. \ref{fig:final_loss_acc_curve}(b), we can observe that ILD-VIT achieves $> 85\%$ accuracy on the validation set. In \textbf{Exp-2:} we randomly split the entire database into 70\%-10\%-20\% training-validation-testing sets and performed five-fold cross-validation whose performance is showed in Fig. \ref{fig:final_loss_acc_curve}(c), which indicates that an average $ACC$, $SNS$, $SPF$, $IS$, and $FS$ of 84.34\%, 83.92\%, 84.45\%, 84.19\%, and 83.67\% is achieved.

\begin{figure}
    \centering
    \includegraphics[width=0.5\textwidth]{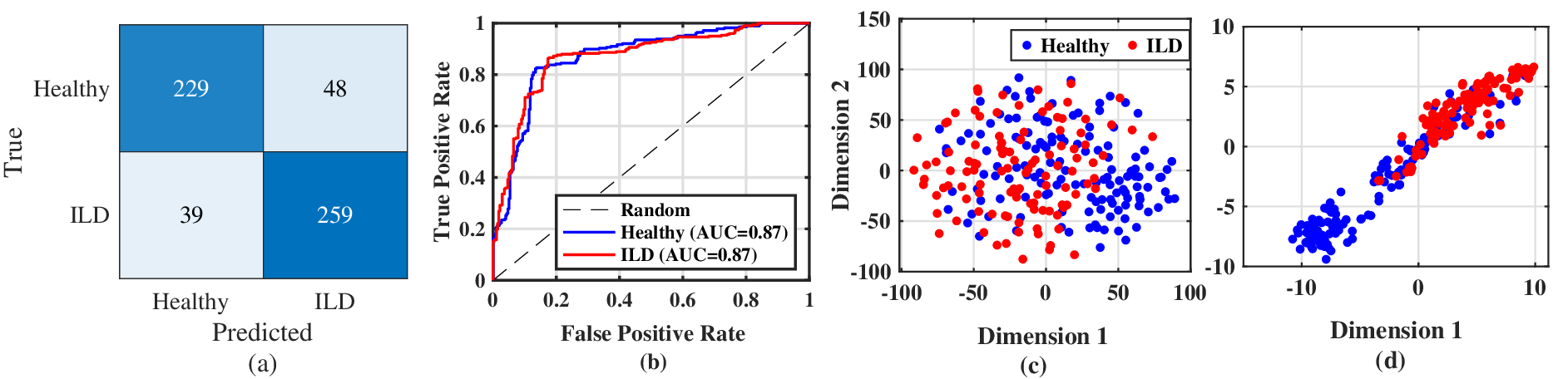}
    \vspace{-0.55cm}
    \caption{Illustrates the (a) confusion matrix and (b) ROC curve, 2D t-SNE visualization of the (c) raw RSs from the test data and (b) their 1D GAP-embeddings from ILD-VIT architecture.}
    \label{fig:t_sne}
\end{figure}

\begin{figure}
    \centering
    \vspace{-0.4cm}
    \includegraphics[width=0.5\textwidth]{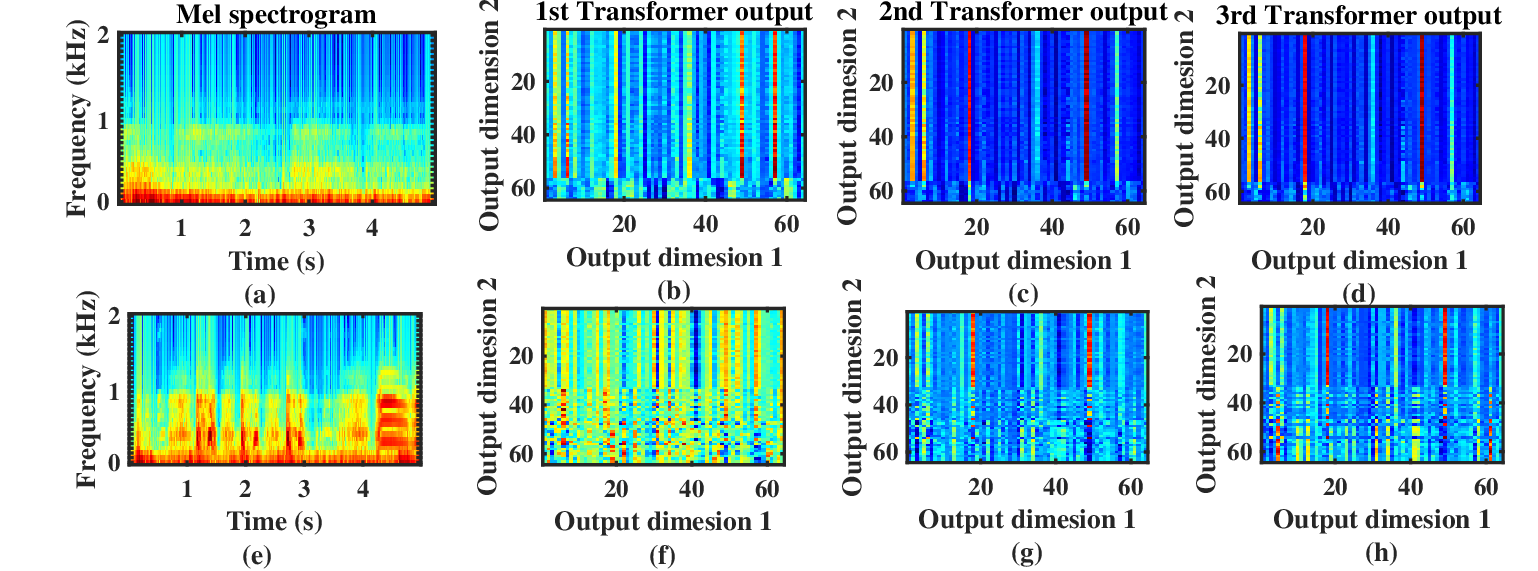}
    \vspace{-0.55cm}
    \caption{Shows the mel spectrogram TFR and activation heatmap visualization of the features extracted from three transformer blocks of ILD-VIT architecture for (a, b-d) healthy and (e, f-h) ILD data.}
    \vspace{-0.4cm}
    \label{fig:heatmaps}
\end{figure}

Additionally, we show the confusion matrix obtained with the test data for \textbf{Exp-1} in Fig. \ref{fig:t_sne}(a), which indicates that a misclassification error rate of 17.32\% and 15.05\% was obtained for healthy and ILD classes. In this case, we have also achieved an $ACC$, $SNS$, $SPF$, $IS$, and $FS$ of 84.86\%, 82.67\%, 86.91\%, 84.79\%, and 84.04\%, after the subject independent blind testing. Further, Fig. \ref{fig:t_sne}(b), shows the receiver operating characteristics (ROC) curve for the ILD and healthy classes, where an area under the curve (AUC) value of 87\% is achieved for both classes. Fig. \ref{fig:t_sne}(c-d) illustrates 2D t-distributed stochastic neighbor embedding (t-SNE)-based feature visualization of the raw RSs and the embedding extracted from the GAP layer for both healthy and ILD classes, respectively. It can be observed that, though initially the embeddings from both classes are randomly scattered in the 2D space in Fig. \ref{fig:t_sne}(c). However, our ILD-VIT, is capable of extracting partially separated class discriminative features as shown in Fig. \ref{fig:t_sne}(d). We have also shown the heatmaps extracted from three transformer blocks of ILD-VIT for correctly classified healthy (Fig. \ref{fig:heatmaps}(a)) and ILD (Fig. \ref{fig:heatmaps}(e)) samples in Fig. \ref{fig:heatmaps}(b-d) and Fig. \ref{fig:heatmaps}(f-h) respectively, which shows that ILD-VIT extract distinctive feature patterns for both the classes.

\begin{figure}
    \centering
    \vspace{-0.4cm}
    \includegraphics[width=0.5\textwidth]{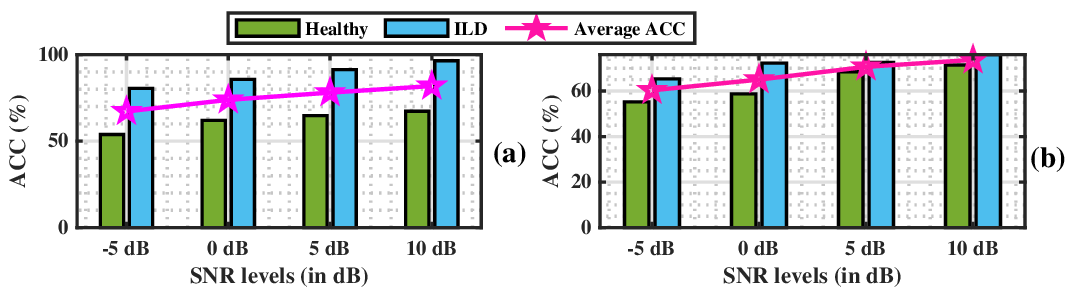}
    \vspace{-0.5cm}
    \caption{Illustrates the class-wise ACC (\%) obtained using ILD-VIT at different SNR levels for Gaussian noise and heart sound noise corrupted cases.}
    \vspace{-0.6cm}
    \label{fig: noise_result}
\end{figure}

We have also experimented the efficacy of the proposed ILD classification system under the influence of various noises such as additive Gaussian noise, and heart sound noise from Physionet-2016 PCG database, under various SNR levels \cite{noise_arka}, and the class-wise $ACC$s under each SNR levels are presented in Fig. \ref{fig: noise_result}(a-b). An overall $ACC$ of 73.54\% and 70.48\% were achieved for ILD classification under the presence of Gaussian noise and heart sounds.

\begin{table}[]
\caption{Overall Performance Comparison of the Proposed ILD-VIT with Recent DL Architecture}
\label{tab:comparison_DL_model}
\resizebox{\columnwidth}{!}{%
\begin{tabular}{|c|c|cccccl|}
\hline
 &  & \multicolumn{6}{c|}{\textbf{Evaluation metrics (\%)}} \\ \cline{3-8} 
\multirow{-2}{*}{\textbf{\begin{tabular}[c]{@{}c@{}}DL \\ architecture\end{tabular}}} & \multirow{-2}{*}{\textbf{\begin{tabular}[c]{@{}c@{}}Trainable\\  parameters\end{tabular}}} & \multicolumn{1}{c|}{\textbf{ACC}} & \multicolumn{1}{c|}{\textbf{SNS}} & \multicolumn{1}{c|}{\textbf{SPF}} & \multicolumn{1}{c|}{\textbf{FS}} & \multicolumn{2}{c|}{\textbf{IS}} \\ \hline
CNN \cite{7422082}& 60,610 & \multicolumn{1}{c|}{69.42} & \multicolumn{1}{c|}{57.48} & \multicolumn{1}{c|}{83.91} & \multicolumn{1}{c|}{67.27} & \multicolumn{2}{c|}{68.14} \\ \hline
\begin{tabular}[c]{@{}c@{}}CNN-LSTM \cite{10345485}\end{tabular} & 1,63,522 & \multicolumn{1}{c|}{71.79} & \multicolumn{1}{c|}{73.84} & \multicolumn{1}{c|}{59.52} & \multicolumn{1}{c|}{71.63} & \multicolumn{2}{c|}{73.03} \\ \hline
SONN \cite{arka_memea} & 74,078 & \multicolumn{1}{c|}{78.63} & \multicolumn{1}{c|}{80.54} & \multicolumn{1}{c|}{80.52} & \multicolumn{1}{c|}{79.86} & \multicolumn{2}{c|}{78.52} \\ \hline
\textbf{\begin{tabular}[c]{@{}c@{}}ILD-VIT\end{tabular}} & \textbf{3,49,506} & \multicolumn{1}{c|}{\textbf{84.34}} & \multicolumn{1}{c|}{\textbf{83.92}} & \multicolumn{1}{c|}{\textbf{84.45}} & \multicolumn{1}{c|}{\textbf{83.67}} & \multicolumn{2}{c|}{\textbf{84.19}} \\ 
\cline{3-7}
& & \multicolumn{1}{c|}{\textbf{84.86}} & \multicolumn{1}{c|}{\textbf{82.67}} & \multicolumn{1}{c|}{\textbf{86.91}} & \multicolumn{1}{c|}{\textbf{84.79}} & \multicolumn{1}{c|}{\textbf{84.04}} \\ \hline
\end{tabular}%
}\vspace{-0.5cm}
\end{table}

\subsection{Performance comparison}
In this subsection, we have compared the performance of ILD-VIT architecture for ILD detection with different SOTA neural network architectures in Table \ref{tab:comparison_DL_model} and existing works based on RSs and other modalities \cite{7422082,Martinez} including CT and HRCT image, in Table \ref{tab:comparison_sota}. 

The experimental results from Table \ref{tab:comparison_sota} show that our proposed ILD-VIT outperforms the only existing work on RS-based ILD detection framework by achieving a high performance for both Exp-1 and Exp-2, which suggests that RSs can be exploited as an alternative diagnostic modality for finding biomarkers to detect ILD condition alongside the image-based gold standard modalities.

\begin{table}[h]
\centering
\caption{Overall Comparison of Our Proposed RS-Based ILD-VIT Framework with Works on ILD Detection Using Different Diagnostics}
\label{tab:comparison_sota}
\Large
\resizebox{\columnwidth}{!}{%
\begin{tabular}{|c|c|c|ccc|}
\hline
\multirow{2}{*}{\textbf{Reference}} & \multirow{2}{*}{\textbf{\begin{tabular}[c]{@{}c@{}}Diagnostic\\ modality\end{tabular}}} & \multirow{2}{*}{\textbf{Methodology}} & \multicolumn{3}{c|}{\textbf{Results (\%)}} \\ \cline{4-6} 
 &  &  & \multicolumn{1}{c|}{\textbf{ACC}} & \multicolumn{1}{c|}{\textbf{SNS}} & \textbf{SPF} \\ \hline
\begin{tabular}[c]{@{}c@{}}Anthimopo\\ ulos et al. \cite{7422082}\end{tabular} & \multirow{2}{*}{\begin{tabular}[c]{@{}c@{}}CT\\ images\end{tabular}} & Five-layer CNN architecture & \multicolumn{1}{c|}{85.50} & \multicolumn{1}{c|}{-} & - \\ \cline{1-1} \cline{3-6} 
\begin{tabular}[c]{@{}c@{}}Martinez\\ et al. \cite{Martinez}\end{tabular} &  & \begin{tabular}[c]{@{}c@{}}Ensemble transfer learning\\ based DL model\end{tabular} & \multicolumn{1}{c|}{82.70} & \multicolumn{1}{c|}{-} & - \\ \hline
\begin{tabular}[c]{@{}c@{}}Vishraj\\ et al. \cite{9725491}\end{tabular} & \begin{tabular}[c]{@{}c@{}}HRCT\\ images\end{tabular} & \begin{tabular}[c]{@{}c@{}}Haralick feature extraction, and\\ classification using RF classifier\end{tabular} & \multicolumn{1}{c|}{85.80} & \multicolumn{1}{c|}{-} & - \\ \hline
\begin{tabular}[c]{@{}c@{}}Roy \\ et al. \cite{arka_ild}\end{tabular} & \multirow{2}{*}{\begin{tabular}[c]{@{}c@{}}RSs\\ (BRACETS)\end{tabular}} & \begin{tabular}[c]{@{}c@{}}Temporal RS driven Sinc\\convolution-based ILDNet\end{tabular} & \multicolumn{1}{c|}{81.25} & \multicolumn{1}{c|}{78.85} & 83.33 \\ \cline{1-1} \cline{3-6} 
\begin{tabular}[c]{@{}c@{}}\textbf{ILD-VIT}\end{tabular} &  & \begin{tabular}[c]{@{}c@{}}\textbf{Mel spectrogram driven VIT}\end{tabular} & \multicolumn{1}{c|}{\textbf{84.34}} & \multicolumn{1}{c|}{\textbf{83.92}} & \multicolumn{1}{c|}{\textbf{84.45}} \\ \cline{4-6}
& & & \multicolumn{1}{c|}{\textbf{84.86}} & \multicolumn{1}{c|}{\textbf{82.67}} & \multicolumn{1}{c|}{\textbf{86.91}} \\
\hline
\end{tabular}%
}
\end{table}

\subsection{On-device implementation}
\begin{figure}
    \centering
    \includegraphics[width=0.45\textwidth, height=3cm]{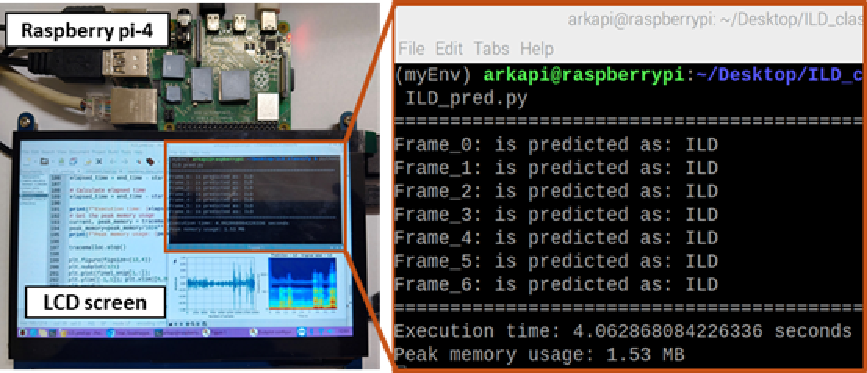}
    \caption{Illustrates the deployment of the proposed ILD-VIT architecture on Raspberry Pi-4 microcontroller.}
    \vspace{-0.35cm}
    \label{fig: implement}
\end{figure}

The proposed framework is implemented on Raspberry-pi-4 microcontroller \cite{10453615} with quad-core ARM Cortex-A7, 1.5 GHz clock frequency, and 8 GB RAM, connected with a LCD touchscreen display as shown in Fig. \ref{fig: implement}. The weight file of the trained ILD-VIT architecture was transferred to the Raspberry-pi, and the relevant Python libraries (Tensorflow, Librosa, Numpy, etc.) were installed to execute the proposed ILD-VIT on the microcontroller. The experimental results show that the proposed work requires a latency of $4.15\pm0.32$ sec and peak memory usage of $1.49\pm0.23$ MB to classify an entire RS of 20 sec length. We also trained and deployed various fine-tuned deep transfer learning models: ResNet50, VGG16, and MobileNetV1 for ILD detection. Table \ref{tab: deployment} demonstrates that our ILD-VIT outperforms other models in terms of inference time, model size, and peak memory utilization on Raspberry-pi.

\begin{table}[h]
\centering
\caption{Performance of ILD-VIT wrt. Other Deep Transfer Learning Networks upon On-Device Deployment using Raspberry-Pi}
\Large
\label{tab: deployment}
\resizebox{\columnwidth}{!}{%
\begin{tabular}{|c|c|c|c|c|c|}
\hline
\textbf{\begin{tabular}[c]{@{}c@{}}DL\\ Model\end{tabular}} & \textbf{\begin{tabular}[c]{@{}c@{}}Model\\ parameters\end{tabular}} & \textbf{\begin{tabular}[c]{@{}c@{}}Model\\ size (MB)\end{tabular}} & \textbf{\begin{tabular}[c]{@{}c@{}}Inference\\ time (sec)\end{tabular}} & \textbf{\begin{tabular}[c]{@{}c@{}}Peak memory\\ (MB)\end{tabular}} & \textbf{\begin{tabular}[c]{@{}c@{}}ACC\\ (\%)\end{tabular}} \\ \hline
VGG16 & 14806940 & 56.48 & 37.91 ± 1.98 & 57.99±1.15 & 51.11 \\ \hline
ResNet-50 & 23910364 & 91.21 & 68.46 ± 1.50 & 97.95 ± 1.01 & 51.99 \\ \hline
MobileNet & 2849436 & 16.92 & 13.02 ± 1.01 & 17.95 ± 0.18 & 59.33 \\ \hline
\textbf{ILD-VIT} & \textbf{349506} & \textbf{1.33} & \textbf{4.15±0.32} & \textbf{1.49 ± 0.23} & \textbf{84.86} \\ \hline
\end{tabular}%
}
\end{table}

\section{Conclusion}
\label{sec: colclusion}
In this letter, we have explored the mel spectrogram TFRs of the RSs with the proposed VIT architecture to identify the ILD condition. By using this approach, we have surpassed the existing RS-based ILD detection work with an $ACC$ of 81.82\%. Further, the successful implementation on the Raspberry-pi microcontroller demonstrates the potential of the proposed framework to be translated into a RS-based standalone medical device for ILD screening. In the future, we intend to investigate various noise elimination techniques to reduce the impact of various metrological factors, such as ambient noise, speech interference, etc., to improve the ILD detection performance.

\bibliographystyle{IEEEtran}
\bibliography{references_list}
\end{document}